# A Tool for the Certification of PLCs based on a Coq Semantics for Sequential Function Charts

Jan Olaf Blech

fortiss GmbH

February 2011

In this report we describe a tool framework for certifying properties of PLCs: CERTPLC. CERTPLC can handle PLC descriptions provided in the Sequential Function Chart (SFC) language of the IEC 61131–3 standard. It provides routines to certify properties of systems by delivering an independently checkable formal system description and proof (called certificate) for the desired properties. We focus on properties that can be described as inductive invariants. System descriptions and certificates are generated and handled using the COQ proof assistant. Our tool framework is used to provide supporting evidence for the safety of embedded systems in the industrial automation domain to third-party authorities.

In this document we describe the tool framework: usage scenarios, the architecture, semantics of PLCs and their realization in COQ, proof generation and the construction of certificates.

## 1 Introduction

Discovering and validating properties of safety critical embedded systems has been a research topic during the last decades. Properties ensuring consistency may be established in several ways. They may either be specified by humans during the development process of an embedded system. In this case they may have the nature of requirements. Another way is the usage of techniques such as model checking and static analysis. *Automatic verification tools* based on model checking and static analysis techniques are used in various software and hardware development projects. Automatic verification tools are successfully applied for a) the discovery of properties and b) the verification of human specified properties providing either some counter-examples or some evidence of correctness, e.g., ensuring that critical parts of systems behave in a specified way. Thus, automatic verification tools increase confidence in the system design.

However, even the verdicts about systems provided by automatic verification tools may be erroneous, since automatic verification tools are likely to contain errors themselves: they use sophisticated algorithms, resulting in complicated implementations. Due to this high level of complexity of their algorithms and the underlying theory, they are hardly ever considered as trustable by certification authorities.

In contrast to general purpose *higher-order theorem provers*, an automatic verification tool possesses a high degree of automation, but it does not achieve the same level of trustability and is usually specialized towards a distinct problem domain. Higher-order theorem provers, like COQ [15], are based on a few deduction rules and come with very small, simple, and trusted proof checkers which are based on type checking algorithms and provide a high level of confidence. For this reason higher-oder theorem provers may be used to re-check properties that have been discovered by automatic verification tools or stated by humans. Even more important, for properties that have been established by humans without using automatic verification tools, the higher-order theorem prover is predestinated to become the only tool to establish a machine check of these properties.

If such a check succeeds one lifts these properties to the higher-level of confidence provided by the higher-order theorem prover. The COQ environment has been accepted by French governmental authorities in a certification to the highest level of assurance of the Common Criteria for Security [14].

Based on our ideas on certification of properties for a modeling language [8, 9] and our work on a certificate generating compiler [6] we present a tool framework CERTPLC which emits certificates and allows reasoning about properties of models for PLCs (Programmable Logic Controller) provided in the SFC (Sequential Function Chart) language. Our work comprises a generation mechanism for COQ representations of our models – a kind of compiler that emits COQ readable files for given models. In addition to this, it comprises other related proof generation mechanisms and a framework for proving automatically that distinct properties of these models do hold. Our COQ certificates – system description, properties and their proofs – are based on an explicit semantics definition of the SFC language, thereby ensuring that correctness conditions hold for distinct systems.

In this report we focus on the following aspects of the CERTPLC tool framework:

- usage scenarios,

- tool architecture,

- semantics of PLCs and their realization in COQ,

- proof generation and the construction of certificates,

- and additional implementation issues.

A detailed description of larger case studies remains subject to future work. Our certification framework is mostly characterized by:

- The usage of an explicit semantics for properties and systems. This is human readable, an important feature to convince certification authorities.

- The focus on the PLC domain and the integration in an existing tool.

- A high degree of automation, that still allows human interaction.

The high expressiveness of our semantics framework is largely facilitated by the usage of a higher-order theorem prover.

## 1.1 Certification

In the context of this paper we define

- *certification* as the process of establishing a certificate.

- *automatic certification* is the process of establishing a certificate automatically.

- In our work *certificates* comprise a formal description of a system, a formal description of a desired property and a proof description that this property does hold.

- *certificate checking* is the process of checking that the property does indeed hold for the formal system description in the certificate. This checking is done by using the proof description in the certificate.

## 1.2 The Trusted Computing Base in Certification

Apart from components like operating system and hardware, in our certification approach, the trusted computing base (TCB) comprises the certificate checker (the core of the COQ theorem prover) and the program that generates formal PLC descriptions for COQ automatically. The check that these descriptions indeed represent the original PLCs can be done manually. One goal for the generation is human readability to make such a check feasible at least for experienced users. Not part of the TCB are the proof description and its generation mechanism. The proof description only provides hints to the certificate checker. In case of faulty proof descriptions a valid property might not be accepted by a certificate checker. It can never occur that a faulty property is accepted even if wrong proof descriptions are used. Thus, our approach is sound, but not necessarily complete.

## 1.3 Related Work

Notable milestones on frameworks to certify properties of systems comprise proof carrying code [20]. Proofs for distinct properties of programs are generated during the compilation of these programs. These are used to certify that these properties do indeed hold for the generated code. Thus, users can execute the certified code and have, e.g., some safety guarantees. At least two problems have been identified:

1. Properties have to be formalized with respect to some kind of semantics. This is sometimes just implicitly defined.

2. Proof checkers can grow to a large size. Nevertheless, they have to be trusted.

The problem of trustable proof checkers is addressed in foundational proof carrying code [2, 27]. Here the trusted computing base is reduced by using relatively small proof checkers. The problem of providing a proof carrying code approach with respect to a mathematically founded semantics is addressed in [23]. In previous work we have also addressed the problem of establishing a formal semantics for related scenarios [6, 9, 5].

Formal treatment of PLCs and the IEC 61131–3 standard has been discussed by a larger number of authors before. Formalization work on the semantics of the Sequential Function Charts is given in [12, 13]. This work was a starting point for our formalization of SFC semantics.

The paper [4] considers the SFC language, too. Untimed SFC models are transformed in to the input language of the Cadence SMV tool. Timed SFC models are transformed into timed automata. These can be analyzed by the Uppaal tool.

Work in the formal treatment of FBDs can be found in [29, 28]. The FBD programs are checked using a model-checking approach.

The approach presented in [25] regards a translation from the IL language to an intermediate representation (SystemC). A SAT instance is generated out of this representation. The correctness of an implementation is guaranteed by equivalence checking with the specification model.

Apart from this, automatic certification of results of tools is an established research field. It has been studied in the contexts of model checkers (e.g., [19, 24, 17]) and SAT and SMT solvers (e.g., [30, 18]). More recently the problem has been regarded in the context of higher-order theorem prover certificates for these tools (e.g., [26, 11, 1]).

### 1.4 Overview

The tool environment in which our CERTPLC tool framework is supposed to be used and an overview about the tool's architecture is described in Section 2. We present the IEC 61131–3 standard, including the SFC language and its semantics as formalized in COQ in Section 3. The CERTPLC ingredients and their interactions are described in some detail in Section 4. Typical proofs that can either be generated or hand-written by using our semantics are discussed in Section 5. Finally, an implementation overview and a short evaluation is given in Section 6. A conclusion is featured in Section 7.

## 2 The Tool Setting

In this section we describe our CERTPLC tool's architecture and usage scenarios. Figure 1 shows the CERTPLC ingredients and their interconnections. In an invocation of the tool framework an SFC model is given to a

- **representation generator** which generates a COQ representation out of it. This is included in one or several files containing the model specific parts of the semantics of the SFC model. The COQ representation is human readable and can be validated against the original graphical SFC specification by experienced users.

The same SFC model is given to a

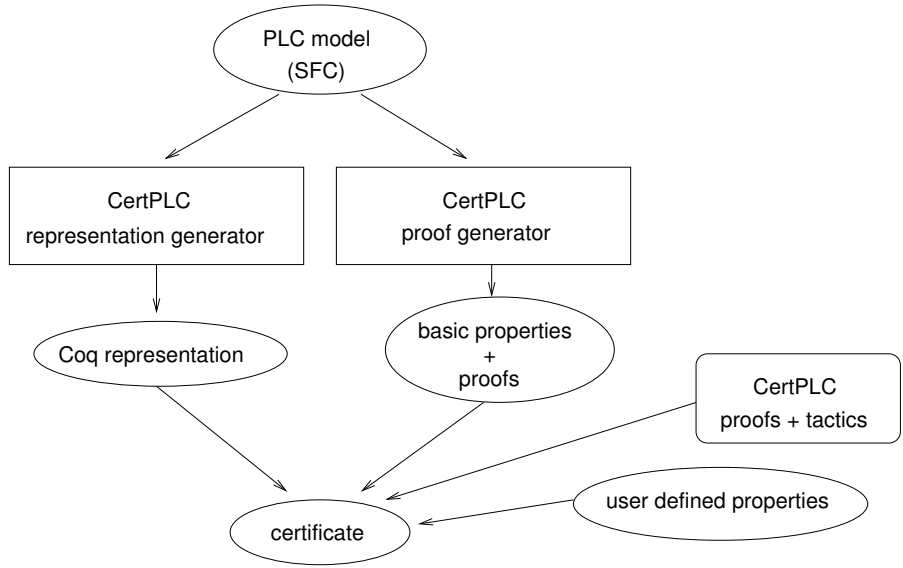

Figure 1: CERTPLC Overview

- **proof generator** which generates COQ proof scripts that contain lemmas and their proofs for some basic properties that state important facts needed for machine handling of the proofs of more advanced properties. For example a proof script is generated for a fact that certain semantical artifacts in all reachable states of the PLC system are only those specified by the syntactic PLC descriptions: The models may have no unspecified behavior.

In order to achieve a certificate one needs a property that the certificate shall ensure. One needs to formalize this desired property. The property is proved in COQ by using, e.g., a provided tactic. This tactic uses the generated properties and their proofs – provided by the proof generator – and a collection of

- **proofs and tactics**, a kind of library. It contaisn additional preproved facts and tactics which may be used to automatically prove a class of properties.

System description, used lemmas and their proofs, and the property and its proof form a certificate.

Furthermore, we have a COQ library that generated and non-generated COQ files can use. It provides often used definitions in addition to the elements described so far. We provide a library of behavioral definitions of PLC blocks which are typically modeled in other languages than SFC. These can be extended by writing the semantics of user defined SFC language functional elements in a functional programming language style. An automatic import for other languages of the IEC 61131–3 standard remains subject to future work.

## Usage Scenario

CERTPLC is developed to support the following standard usage scenario:

- A PLC is developed starting with the following steps:
    1. Establishing requirements,
    2. and derive some early formal specification.
    3. Based on this specification the overall structure – e.g., the control flow – is specified using the SFC language. More fine-grained behavioral aspects are textually specified, e.g., by annotating the SFC structure.
    4. Taking the requirements and this specification, developers using the help of automatic verification tools derive and specify consistency conditions and properties that must hold. Some consistency conditions may directly correspond to a subset of the requirements.
- Regarding 3) the SFC structure is modeled in the graphical EasyLab tool [3] or imported into EasyLab.
- Regarding 4) properties and SFC action blocks are specified using the COQ syntax by trained developers. It is not required to do any proofs in COQ for this.
- CERTPLC generates representations for the PLC specification. Together with the properties a certificate is established automatically or with minor user interaction: the choice of tactics.
- The PLC development is further refined and fine grained parts may be implemented using other languages from the IEC 61131–3 standard.
- Certificates may be either regenerated – if possible – or manually adapted – in case of unsupported language elements that may occur during the refinement – to cope with possible changes.

The certificate can be distributed and analyzed independently by third parties. One overall goal is to convince certification authorities and potential customers of the correctness of PLCs with the help of certificates. Since the certificates are independent of the original development and its tools some confidential data does not have to be revealed during the process of convincing customers or certification authorities.

The described usage scenario can be adapted. It is, e.g., possible to integrate hand written specifications and proofs.

## 3 IEC 61131–3, SFCs, their Semantics and Certification

In this section we sketch the semantics of Sequential Function Charts (SFCs), partially the semantics of Function Block Diagrams (FBDs). We sketch their realization in the COQ environment. The description in this Section is based on our earlier work [7] which is influenced by the descriptions in [12, 13].

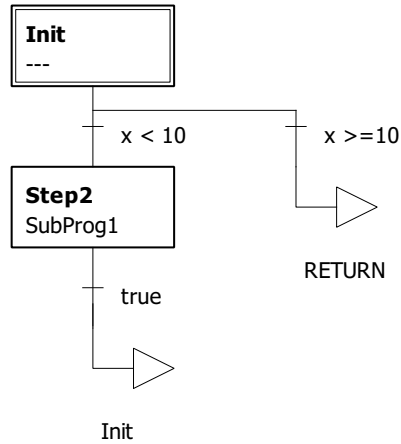

Figure 2: A Loop in the SFC language

## 3.1 The SFC Language

Our tool framework works with PLCs described in the SFC language. The SFC language is a graphical language for modeling PLCs. It is part of the IEC 61131–3 standard [21] and frequently used together with other languages of this standard. In such a case, SFCs are used to describe the overall control flow structure of a system. The standard is mainly used in the development of embedded systems in the industrial automation domain.

The standard leaves a few semantical aspects open to the implementation of the PLC modeling and code generation tool. In cases where the semantics is not well defined by the standard we have adapted our tool to be compatible with the EasyLab [3] tool.

**Syntax** Syntactically we represent an SFC as a tuple $(S, S_0, T, A, F, V, Val_V)$. It comprises a set of steps $S$ and a set of transitions $T$ between them. A step is a system location which may either be active or inactive in an actual system state, it can be associated with SFC action blocks from a set $A$. These perform sets of operations and can be regarded as functors that update functions representing memory. Memory is represented by a function from a set of variables $V$ to a set of their possible values $Val_V$. The mapping of steps to sets of action blocks is done by the function $F$.

A transition is a tuple $(S_{in}, g, S_{out})$. It features a set of states that have to be enabled $S_{in} \subseteq S$ in order to take the transition. It features a guard $g$ that has to be evaluated to true for the given system state. $g$ is a function from system memory to a truth value – in COQ we formalize this as a function to the *Prop* datatype. A transition may have have multiple successor steps $S_{out} \subseteq S$. The types $Val_V$ that are formalized in our SFC language comprise different integer types and boolean values.

$S_0 \subseteq S$ is the set of initially active steps.

Figure 2 shows an example SFC structure realizing a loop with a conditional branch. The execution starts with an initialization step *init*. After it has been processed control may pass to either *Step2* or to a step *Return*. Once *Step2* has been processed control is passed to *init* again.

Please note, that in addition to loops and branches SFCs allow for the formalization of parallel processing and synchronization of control. This is due to the multiple successor and predecessor steps in a transition.

The COQ realization of the SFC syntax follows the presented description. For compatibility with the EasyLab tool and to ease generation we distinguish between steps and step identifiers in our COQ files, thereby introducing some level of indirection.

**Semantics** Semantically the execution of an SFC encounters states, which are $(m, a, s)$ tuples. They are characterized by a memory state $m$, the function from variables to their values, a set of active action blocks $a$ that need to be processed and a set of active steps $s$.

The semantics is defined by a state transition system which comprises two kinds of rules:

1. A rule for processing of an action block from the set of active action blocks $a$. This corresponds to updating the memory state and removing the processed action block from $a$.

2. A rule for performing a state transition. The effect on the system state is that some steps are deactivated, others are activated. Each transition needs a guard that can be evaluated to true and a set of active steps. Furthermore, we require that all pending action blocks of a step that is to be deactivated have been executed.

It is custom to specify the timing behavior of a step by time slices: a (maximal) execution time associated with it. In our work, this is realized using special variables that represent time.

## 3.2 The FBD Language

Function block diagrams are a language from the IEC 61131–3 used to model the behavior of action blocks in SFCs. Other languages that may be used for this purpose comprise instruction lists (ILs) and ladder diagrams (LDs).

FBDs comprise two basic kinds of elements: function blocks and connections between them. Each function block represents an instruction. There are special instructions for reading and writing global variables. Edges between function blocks are used to model dataflow. Thus, FBDs are used to describe functions.

Apart from the basic functionality, FBDs may contain cycles in their dataflow description. Semantically such a cycle must feature a delay element. Variable values associated with such an FBD are computed in an iterative process.

In the case of cyclic dependencies an FBD has to be associated with a time slice, a maximal time – number of iterations – for the execution of the FBD. Thus, on an abstract level, FBDs may still be regarded as functions and as SFC action blocks.

We have formalized an FBD syntax and semantics framework in COQ that follows the description above. Most parts of this, however, are only to be used manually by users who manually change system descriptions and corresponding proofs.

# 4 The CertPLC Tool Environment and Coq

In this section we describe the CERTPLC tool. We describe its architecture and implementation. Furthermore, we describe the COQ components. We present some static COQ code that is generic to our framework. Furthermore, we present some PLC specific example COQ code – definitions and proofs – to demonstrate aspects of its generation.

## 4.1 Starting Point: Semantics of SFCs in Coq

Taking the semantics sketch of SFCs in Section 3 the semantic representation of the SFC structure is encoded in COQ as a kind of transition system. For each given SFC $SFC$ we generate a COQ representation. It specifies a set of reachable states and a transition relation. The transition relation is a set of tuples:

$$\text{preceding state} \times \text{succeeding state}$$

It is denoted $[\![SFC]\!]$.
The set of reachable states for $SFC$ is denoted $Reachable_{SFC}$.

## 4.2 Realization Using Generic and Generated Files

In order to certify properties of PLCs we need files that contain semantics of systems, interesting properties and proofs of these properties. Some of these files are generic, i.e., they can be used for a large class of PLCs, properties, and proofs. We provide a library of static files that contain generic aspects. Other files are highly specific to distinct PLCs. For each PLC CERTPLC generates files that are just needed for this particular PLC, properties formulated on it, and proofs that can be conducted on it.

In particular the following aspects are generic, thus, stored in static files:

- Generic definitions and templates for SFCs:
    - Datatype definitions and derived properties of these datatypes.
    - Definitions for building blocks: SFC action blocks, FBD blocks, and common combinations of these blocks.
    - Generic semantics framework comprising an instantiable state transition relation and a generic definition for a set of reachable states.
- Tactics for solving certain proof aspects:
    - Tactics that contain an overall proof structure, deal with certain system structures and property structures.
    - Tactics that solve arithmetic constraints.

The following aspects are individual for each PLC, thus, they are generated:

- A state transition like representation of SFCs formalized using generic SFC definitions and a concrete definition of reachable states instantiating a generic SFC definition.

- Lemmas containing distinct facts on the PLC and their proofs.

Furthermore, the properties that shall hold are of course specific to each PLC. Their verification is done by either using a tactic that assembles the generic and non-generic parts of the proof or by some hand-written proof script adaptations.

## 4.3 Generic / Static parts of the Coq Infrastructure

Here we describe generic parts of the COQ parts in our CERTPLC tool framework. These are realized as static COQ files and can used by the dynamically created files.

**Datatypes** Datatypes which we have formalized for SFCs comprise integers of different length (8,16,32 bit) and bools. In COQ they are stored using the datatype *nat* of natural numbers plus a flag that tags them as being member of a distinct integer types. Operators working on these integers perform operations compliant with the type. Other datatypes like floating point values seem not so important for most PLC applications and are not yet supported, although they could be integrated relatively easy: The basic semantics definitions in our framework are able to support a much richer type system, even dependent datatypes.

**Building Blocks** Building blocks define common elements for the construction of PLCs. Two levels of building blocks can be distinguished:

- Function blocks that are intended to become part of FBDs.

- Predefined action blocks. These may be, but do not have to modeled using FBDs.

As mentioned in Section 2 we have formalized a library of these blocks which can be extended. Extending the library is usually done when considering new case studies since different application domains have different sets of FBD and SFC elements. FBD elements that are highly specific to a distinct application are highly common in PLCs. For FBDs we have experienced even vendor specific elements for the basic arithmetic operations.

**Generic Semantics Framework** Most importantly, our semantics framework comprises a template for a state transition relation of PLC systems and a template for defining the set of reachable states. In order to realize this, we first define generic instantiable predicates that formalize a state transition relation. We provide a predicate *executeAction* defined in Figure 3. It formalizes the effect of the execution of an action block: The predicate takes two states (sometimes called configurations $c$ and $c'$) and returns a value of type *Prop*. We require four conditions to hold in order to take a state transition:

1. An action block $a$ needs to be active.

2. The memory mapping after the transition is the application of $a$ to the previous memory mapping. This is the updating of the memory by executing the action block.

3. The action block $a$ is removed from the set of active action blocks during the transition.

```
Definition executeAction:
   ((X -> D X) * list Sid  * list A) ->  ((X -> D X) *  list Sid * list A)
     -> Prop :=
fun c c' =>
   let '(f,activeS,activeA) := c in
   let '(f',activeS',activeA') := c' in
   (exists a, In a activeA /\ f' = a f /\
    activeA' = remove Action_eq_dec a activeA) /\ activeS = activeS'.
```

Figure 3: The *executeAction* Predicate

4. The rest of the state does not change.

The *stepTransition* predicate defined in Figure 4 formalizes the effect of a transition from an SFC step to another. Here we require the following items:

1. The validity of the transition (first three conjunct subexpressions).

2. The memory state may not change.

3. The activation of steps is semantics conform.

4. The activation and requirements of action blocks is semantics conform.

The figure shows another disjoint part which corresponds to the reactivation of an action block belonging to an active step. This is a semantics feature which is needed by some operation modes of PLCs. Using these predicates we define inductively the set of reachable states in Figure 5. This predicate depends on

- an initial state (comprising a list of initially active steps),

- and a transition relation.

It is defined as described in Section 3. The definition uses two rules stating that

- the initial state is in the set of reachable states (Rule *SFCrsp_init*) and

- the condition under which a succeeding state to a given state is in the set of reachable states (Rule *SFCresp_transrule*).

The second rule makes use of the two predicates defined above.

Apart from the version presented here, we have defined a version that uses priorities to resolve non-determinism in the choice of action block activation and transitions. This is may be advantageous to cope with some implementation issues raised by PLC development tools.

```
Definition stepTransition
   (transitions: list (list Sid * ((X -> D X) -> B) * list Sid)) :
     ((X -> D X) * (list  (Sid)) * list A) ->
     ((X -> D X) * (list   (Sid)) * list A)
   -> Prop :=
fun c c' =>
   let '(f,activeS,activeA) := c in
   let '(f',activeS',activeA') := c' in
      (exists t, let '(t_src,g,t_tgt):=t in
        In t transitions  /\ incl t_src activeS /\
        Is_true (interpretB (g f)) /\ f = f' /\
        activeS' = (filter
          (fun x => negb (existsb
             (fun x' =>
                if (Sid_eq_dec x x') then true else false)
            t_src)) activeS) ++ t_tgt /\
        Is_true (forallb
         (fun a => (negb (existsb
            (fun x => if Action_eq_dec a x then true else false )
           activeA))) ( fold_left (fun l x => l ++ Sid2A x) t_src nil)) /\
        activeA' =
          (fold_left (fun l x => l ++ Sid2A x) t_tgt nil) ++ activeA)
   \/ (
   f = f' /\ activeS' = activeS /\
   (exists s, In s activeS /\
    Is_true (forallb
      (fun t => let '(t_src,g,t_tgt):=t in
        if (existsb
            (fun s' =>  if (Sid_eq_dec s s') then true else false)
             t_src) then negb (interpretB (g f)) else true)
      transitions)  /\
    activeA' = Sid2A s ++ activeA)).
```

Figure 4: The *stepTransition* Predicate

```
Inductive SFCreachablestates_pred
   (init: (X -> D X) * list Sid * list A)
   (transitions: list (list Sid * ((X -> D X) -> B) * list Sid)) :
        ((X -> D X) * list Sid * list A) -> Prop :=
| SFCrsp_init :
   SFCreachablestates_pred
      init transitions init
| SFCresp_transrule: forall c c',
    SFCreachablestates_pred
       init transitions c ->
    (executeAction c c' \/ stepTransition transitions c c' ) ->
    SFCreachablestates_pred
      init transitions c'.
```

Figure 5: Formalization of Reachable States

**Structural Tactics** We have established tactics that perform the most basic operations for proofs of properties. They work with semantics definitions based on our generic semantics framework. Depending on the property such a tactic is selected by the user and applied as the first step in order to prove the desired property. Different tactics have to be selected by the user: Selection depends on whether the property is some kind of inductive invariant – the default case mostly addressed in this paper- , or another distinct class of properties. We have identified several other classes that are relevant for our case studies. Such a tactic is applied as the first step in order to prove the desired property. These tactics already perform most operations concerning the system structure. Especially for the non-standard cases, tactics applications may leave several subgoals open. These may be handled with more specialized tactics tailored for the corresponding characteristics of these proof-goals.

We are able to prove a large class of commonly appearing inductive properties with a single default tactics application.

**Arithmetic tactics** Arithmetics tactics solve subgoals that appear at later stages in the proof. They may be called by structural tactics or work on open subgoals that are left open by these tactics. They comprise classical decision procedures like (e.g., Omega [22] – its implementation in COQ).

In addition to structural tactics and arithmetic tactics we are also working on tactics that combine arithmetic aspects with other system state dependent information.

## 4.4 Semantics Definitions as State Transition Systems

As seen in Section 4.3 we only need to instantiate a template in order to create a system definition that captures the semantics of our PLCs. We need to provide at least a set of initially active steps, a transition relation, and action block definitions.

For the initial step, we provide an initial memory state, where all values are set to a default value and a single active entry step.

The transition relation is generated by translating the SFC transition conditions into COQ. The generated COQ elements of the transition relation for the SFC depicted in Figure 2 are shown in Figure 6. Three tuples are shown, each one comprises a set of activated source steps, a condition and a set of target steps activated after the transition. It can be seen that the condition maps a variable value mapping – part of the SFC state – to a truth condition – returning the type *Prop*. The types used in this expression are 16-bit integer types.

Appropriate action blocks are selected by their names. In addition, to this, we generate several abstract datatype definitions for identifying steps with names and identifiers and function blocks and action blocks.

## 4.5 Automatically Generated Proofs for Distinct Facts

Some proofs are automatically generated for each system. These prove distinct basic facts of the system. These proofs are used automatically by tactics, but can also be used manually to prove user defined properties of systems.

```
( Init::nil ,
  fun f => ((fun  (x : int16) => x <int16 10 )  (f VAR_x) ),
  Step2::nil )

( Step2::nil ,
  fun f => ((fun  (x : int16) => 1 )  (f VAR_x) ),
  Init::nil )

( Init::nil ,
  fun f => ((fun  (x : int16) => x >=int16 10 )  (f VAR_x) ),
  Return::nil )
```

Figure 6: Generated Transition Rules in COQ

```
Lemma aux_1:
   forall s, SFCreachable_states s -> (forall a, In a (snd s) ->
      ( a = action_Init \/  a = action_Step1 )).
```

Figure 7: An Automatically Generated Basic Property

One important fact that needs to be proven is that only those action blocks may appear in the set of currently active action blocks that do belong to the actual system definition. Our proof generator generates an individual lemma and its proof for each PLC. Figure 7 shows such a lemma for an SFC that comprises just two possible action blocks: *action_Init* and *action_Step1*. The predicate *SFCreachable_states* is created by instantiating the generic definition from Figure 4 for a concrete PLC. *In* and *snd* (second) are COQ functions to denote membership in a set and select an element of a tuple, respectively. In the case at hand *snd* selects the set of active action blocks from a state. The proof itself is also generated. It comprises an induction on reachable states of the concrete system. To give a look and feel, the complete generated proof script for the Figure 7 is shown in Figure 8.

### 4.6 Formalizing Properties of Systems

The certification of properties is the key feature of CERTPLC. Up till now, users have to write their desired properties in COQ syntax. This does not require as much understanding of the COQ environment as one could think at a first glance. All that is required is writing a logical formula that captures the desired property.

A very simple invariant property for a concrete SFC system is depicted in Figure 9. One can see that a certain condition has to hold for all reachable states $s$ of a system. Again, the predicate *SFCreachable_states* is created by instantiating the generic definition from Figure 4. The condition states that the value of a variable *Y_VAR* has to be larger than $0$ in every state encountered in the system. *Is_true* and *SFCvartype2bool* are functions to map the truth value of this expression to the *Prop* type provided by COQ.

```
intros x P.
eapply SFCreachablestates_pred_ind with (p:=x) (init:= SFCinitstate).
simpl; intros; contradiction.
Focus 2.
apply P.
intros c c'.
intros HA HB HC.
destruct HC as [HC  | HC].
unfold executeAction in HC.
destruct c as ((c1,c2),c3).
destruct c' as ((c1',c2'),c3').
intros a'.
destruct HC as [a HC].
unfold snd.
intros HD.
unfold snd in HB.
pose (HBa := (HB a'));case (ActionDecEq a' a);intros Aeq.
rewrite <- Aeq in HC.
destruct HC as [HC1 [HC2 HC3]].
apply HBa in HC1; assumption.
destruct HC as [HC1 [HC2 HC3]].
rewrite HC3 in HD.
apply ElementRemover in HD.
apply HBa in HD.
assumption.
assumption.
unfold stepTransition in HC.
destruct c as ((c1,c2),c3).
destruct c' as ((c1',c2'),c3').
unfold snd.
intros a HD.
destruct HC as [HC | HC].
destruct HC as [[[t1 t2] t3]  HC].
destruct HC as [HC1 [HC2 [HC3 [HC4 [HC5 [ HC6 HC7]]]]]].
rewrite HC7 in HD.
simpl in HC1.
or_destruct HC1;
  try (apply snd_prop in HC1; unfold snd in HC1; rewrite <- HC1 in HD; simpl in HD;
    auto; destruct HD); try contradiction; (auto).
destruct HC as [HC1 [HC2 HC3] ];
destruct HC3 as [s' [HC3a [HC3b HC3c] ]];
destruct s';
simpl in HC3c;
rewrite HC3c in HD;
simpl in HD;
destruct HD; auto.
Qed.
```

Figure 8: An Automatically Generated Proof Script

```
Lemma generic_proof_simple_condition:
  forall s,
    SFCreachable_states s ->
    Is_true
      (SFCvartype2bool (0 <int16 (SFCVarMap s) Y_VAR)) .
```

Figure 9: A Simple Invariant

## 5 Automatic Certification of Invariant Properties

In this section we describe the principles of automatically proving properties correct. Proof scripts encapsulating these principles are generated by the CERTPLC framework components as described in Section 4.

Two sources of properties can be distinguished:

- Our proof generator generates lemmas and their proofs stating basic facts. These are useful for proving inductive invariants and additional properties, e.g., consistency conditions.

- Additionally we have designed and reused tactics which are able to prove inductive invariants containing classes of arithmetic expressions (e.g., Presburger arithmetics in invariants, action blocks, and guard expressions) automatically.

The process of proof generation follows our work described in [8, 9]. In contrast to [8, 9] we split the process of proof generation into a system specific part which is generic with respect to the property and general tactics to prove an inductive invariant containing arithmetic expressions.

### 5.1 Proof Structure for Inductive Properties

We start with an inductive invariant property $I$ and an SFC description of a PLC $SFC$. Following the ideas presented in [9] the structure of a proof contained in our certificates is realized by generated proof scripts, generic lemmas and tactics. They establish a proof principle that proves the following goal:

$$\forall s \,.\, s \in Reachable_{SFC} \implies I(s)$$

$Reachable_{SFC}$ is the inductively defined set of reachable states, $[\![SFC]\!]$ specifies the state transition relation (cf. Section 4). First we perform an induction using the induction rule of the set of reachable states. This rule is automatically established by COQ when defining inductive sets. After the application the following subgoals are left open:

1. $I(s_0)$ for initial states $s_0$,

2. $I(s) \wedge (s, s') \in [\![SFC]\!] \implies I(s')$

The first goal can be solved by some relatively simple tactic which just checks that all conditions derived from $I$ are fulfilled in the initial states.

For the second goal the certificate realizes a proof script which – in order to allow efficient certificate checking – performs most importantly the following operations:

- Splitting of conjunctions in invariants into independently verifiable invariants.

- Splitting of disjunctions in invariants into two independently verifiable subgoals.

- Normalizing arithmetic expressions and expressions that make distinctions on active steps in the SFC.

- Exhaustive case distinctions on possible transitions. Each case distinction corresponds to one transition in the control flow graph of the SFC. A typical case can have the following form:

    Precondition on states associated with a case distinction
    Transition condition associated with a case distinction
    Conditions on possible reachable states after one transition
    $\Longrightarrow$
    Property holds for succeeding states

    The elements in such a goal can, e.g., feature arithmetic constraints, which can be split into further cases.

    Some of the cases that occur can have contradictions in the hypothesis. For example one can imagine an arithmetic constraint for a variable from a precondition of a state contradicting with a condition on a transition. These contradictions result from the fine granularity of our case distinctions. Some effort can be spend to eliminate contradicting cases as soon as possible (cf. [9]) which can speed up the checking process.

    Unlike in classical model-checking we get the abstraction from (possibly infinite) concrete states to (finite) arcs in the control flow graph almost for free.

- The final step comprises the derivation of the fact that the invariant holds after the transition from the transition conditions and the decision of possible arithmetic constraints.

### 5.2 Proving Non Arithmetic Invariants

The main focus of CERTPLC is on inductive invariants, However, additionally we have established a library of preproved lemmas useful for proving the (un-)reachability of certain states. In particular the following cases turned out to be necessary in our case studies:

- State $s$ can only be reached via a transition where a condition $e$ must be enabled, $s$ is not initial, $e$ can never be true in the system, this implies $s$ can not be reached.

- Under system specific preconditions: Given an expression over states $e$, if $e$ becomes true the succeeding state will always be $s$. This is one of the few non-inductive properties. However, the proof of this benefits from a proof that $e$ can only become true in a number of distinct states. This can be provided by one of the techniques above.

Additional consistency properties may be certified by hand-written proof scripts. This, however, requires some level of expertise in COQ.

# 6 Additional Implementation Aspects and Evaluation

Here we describe additional implementation aspects that are not covered in the previous sections and provide a short evaluation.

**Additional Implementation Aspects** The COQ representation generator is implemented as an Eclipse plug-in in Java using the IEC 61131–3 meta model of EasyLab and the Eclipse Modeling Framework (EMF) [16]. Representations and lemmas + proofs for basic properties are generated for COQ 8.3. Likewise our libraries for tactics, lemmas and SFC action blocks are formalized using this version. The realization of this representation generator can be regarded as a simple compiler or model to model transformation. A kind of visitor pattern is used to pass through the model representation in EMF format and emit corresponding COQ code. The generation of PLC specific lemmata and their proofs is similar to code generation. A visitor picks all necessary information and generates the lemma text and its proof script. Some storage of intermediate information is needed. The setup is similar to the techniques used in [9] and [10].

Likewise our work builds upon the PLC semantics of EasyLab which we have formally described [7] and realized in COQ.

**Evaluation** We have applied our tool to a number of case studies. Most notably we have verified some consistency properties appearing in a model describing a conveyor belt controller that is part of the EU Artemis ACROSS project.

Completeness of our approach is not targeted in this work so far.

# 7 Conclusion and Future Work

In this report we have presented the CERTPLC environment for certification of PLCs. We have described possible usage scenarios, the technical realization and parts of the COQ semantics. CERTPLC is aimed at the formal certification of PLC descriptions in the SFC language. Nevertheless, some features of FBDs are integrated. Future work shall extend this support and aims at integrating other languages from the IEC 61131–3 standard. A more detailed description of our case studies – as far as possible, without revealing confidential information – is a short term goal. Additional case studies and the collection of experiences of the usage of our tool is another short term goal.

A long term goal could also be the realization of a certification framework for state transition based languages. Further future work could extend the power properties that can be certified and establish completeness proofs for distinct classes of properties.

## Acknowledgements

This work has been supported by the European research project ACROSS under the Grant Agreement ARTEMIS-2009-1-100208.